\documentclass{article}

\usepackage[margin=1in]{geometry} 
\usepackage{amsmath,amsthm,amssymb,hyperref}
\usepackage[utf8]{inputenc} 
\usepackage[T1]{fontenc}    
\usepackage{hyperref}       
\usepackage{url}            
\usepackage{booktabs}       
\usepackage{amsfonts}       
\usepackage{nicefrac}       
\usepackage{microtype}      
\usepackage{graphicx}       
\usepackage[backend=biber, sorting=none, style=numeric-comp]{biblatex}
\usepackage{textgreek}

\usepackage{titlesec}

\titleformat{\section}
  {\normalfont\fontsize{11}{15}\bfseries}{\thesection}{1em}{}
\titleformat{\subsection}
  {\normalfont\fontsize{11}{15}\itshape}{\thesubsection}{1em}{}
  
\usepackage{subcaption}
\DeclareCaptionFormat{custom}
{%
    \textbf{#1#2}\textit{\small #3}
}
\date{}

\begin{document}

\noindent \Large{\textbf{Sensitive multi-species photoacoustic gas detection based on mid-infrared supercontinuum source and miniature multipass cell}}

\normalsize{}

\vspace{5mm}

\noindent \large{\textbf{Tommi Mikkonen,$^1$ Tuomas Hieta,$^2$ Goëry Genty,$^1$ and Juha Toivonen$^{1,*}$ }}
\vspace{3mm}

\noindent \small{$^1$Photonics Laboratory, Physics Unit, Tampere University, FI-33014 Tampere, Finland	

\noindent $^2$Gasera Ltd., Lemminkäisenkatu 59, FI-20520 Turku, Finland	      

\noindent Corresponding author: *juha.toivonen@tuni.fi}

\normalsize{}

\vspace{5mm}

\noindent \textbf{Abstract:} We report multipass broadband photoacoustic spectroscopy of trace gases in the mid-infrared. The measurement principle of the sensor relies on supercontinuum-based Fourier transform photoacoustic spectroscopy (FT-PAS), in which a scanning interferometer modulates the intensity of a mid-infrared supercontinuum light source and a cantilever microphone is employed for sensitive photoacoustic detection. With a custom-built external Herriott cell, the supercontinuum beam propagates ten times through a miniature and acoustically non-resonant gas cell. The performance of the FT-PAS system is demonstrated by measuring the fundamental C--H stretch bands of various hydrocarbons. A noise equivalent detection limit of 11~ppb is obtained for methane (40~s averaging time, 15~\textmu W/cm$^{-1}$ incident power spectral density, 4~cm$^{-1}$ resolution), which is an improvement by a factor of 12 compared to the best previous FT-PAS systems. High linearity and good stability of the sensor provide reliable identification of individual species from a gas mixture with strong spectral overlap, laying the foundation for sensitive and selective multi-species detection in small sample volumes.


\section{Introduction}
Photoacoustic spectroscopy (PAS) is an extensively used technique for trace gas sensing due its high sensitivity and low gas consumption \cite{harren2006photoacoustic,bozoki2011photoacoustic}. These properties result from a unique operation principle converting the detection from optical into acoustic domain. Modulated incident radiation induces periodic absorption in a closed gas cell, which leads to periodic pressure variations via molecular thermal relaxation. The pressure waves proportional to the gas concentration are typically detected with a capacitive microphone. More recently, detection schemes utilizing pressure transducers with enhanced sensitivity such as piezoelectric quartz tuning forks and optically read cantilevers have been demonstrated \cite{patimisco2014quartz,wilcken2003optimization,tomberg2018sub}. The detection of multiple species with high selectivity requires a spectrally broadband system that can be implemented using a Fourier transform photoacoustic spectroscopy (FT-PAS) approach \cite{busse78,farrow1978fourier}. In FT-PAS, a scanning interferometer modulates each wavenumber of a broadband light source at a distinct acoustic frequency, after which all the pressure waves at these frequencies are simultaneously detected in an acoustically non-resonant gas cell. Broadband thermal emitters are typically used as the light source in FT-PAS, but recently supercontinuum (SC) sources and frequency combs have been shown to yield significantly better performance \cite{mikkonen2018broadband,mikkonen2020noise,sadiek2018optical,karhu2019broadband,wildi2020photo}.

The spatial coherence of these light sources enables efficient coupling to an optical cavity, which amplifies optical power inside the gas cell resulting in a proportional enhancement in the photoacoustic (PA) signal and sensitivity. Although non-resonant multipass cavities provide less amplification than resonant cavities \cite{borri2014intracavity,tomberg2019cavity}, they are often preferred due to simpler operation, better mid-infrared (mid-IR) availability and wider spectral bandwidth, which is especially crucial for broadband systems. In single-frequency PAS, various types of multipass cavities have been reported, including simple zigzag configurations \cite{miklos2006multipass,saarela2010mp,chen2019mp,wildi2020photo}, circular cells \cite{manninen2012mp}, retroreflectors \cite{ma2019mp} and Herriott cells \cite{hornberger1995mp,nagele2000mp,nagele2002mp,krzempek2018mp,zhang2020mp2} that are especially attractive for conventional cylindrical gas cells. The reflectors in most miniature multipass systems are internal \cite{miklos2006multipass,saarela2010mp,chen2019mp,manninen2012mp,zhang2020mp2}, i.e. in contact with the gas medium, to eliminate optical power losses from windows, and the background signal from these mirrors is minimized by ways of wavelength modulation \cite{chen2019mp,manninen2012mp,zhang2020mp2} or acoustic resonances \cite{miklos2006multipass,saarela2010mp}, which however are not applicable to FT-PAS.

Here, we report the first broadband (spectral bandwidth over 100~cm$^{-1}$) multipass PAS system in the mid-IR by employing an external Herriott-type mirror configuration. Radiation from a mid-IR SC source is guided ten times through a miniature gas cell equipped with a cantilever microphone that has been shown to perform exceptionally well in broadband PA detection due to its sensitivity over a wide frequency range \cite{hirschmann}. The multipass system provides a sixfold signal enhancement in a bandwidth of 400~cm$^{-1}$ around 3000~cm$^{-1}$, which results in an order of magnitude improvement in sensitivity for FT-PAS systems and detection limits for hydrocarbons in the low parts per billion (ppb) level. We also study the stability of the system and demonstrate its high selectivity showing great promise for multi-species trace gas detection.

\section{Experimental}

\begin{figure}[h]
\centering
  \includegraphics[height=5.7cm]{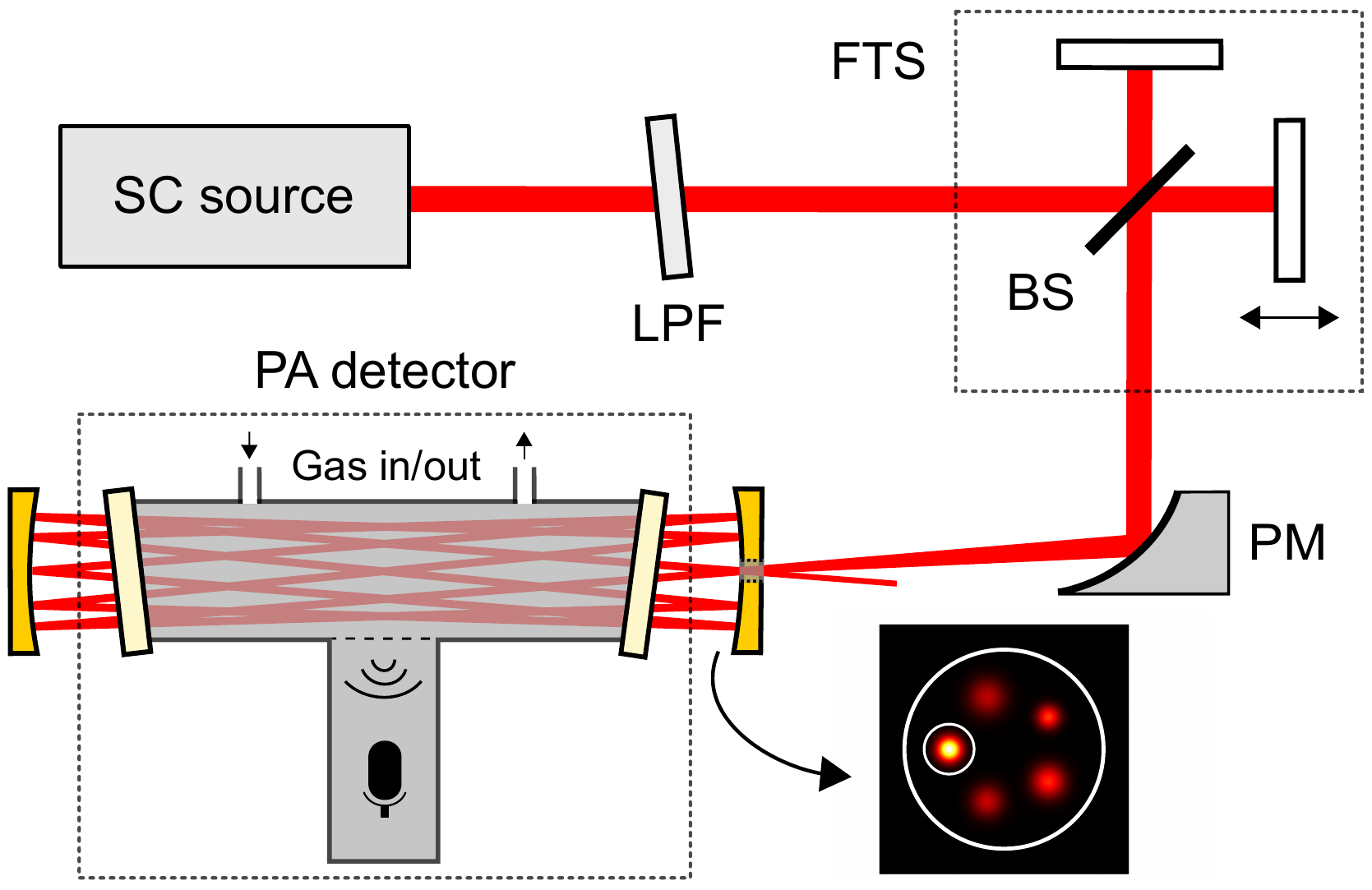}
  \caption{A schematic of the experimental setup: SC - supercontinuum; LPF - long-pass filter; FTS - Fourier transform spectrometer; BS - beam splitter; PM - parabolic mirror; PA - photoacoustic. The simulation on the bottom-right corner shows the theoretical intensity pattern at the front mirror. The smaller white ring represents the input aperture and the larger ring illustrates the gas cell.}
  \label{fig:setup}
\end{figure}

The experimental setup is illustrated in Fig. \ref{fig:setup}. A home-built SC source is generated by injecting sub-nanosecond pulses with 70~kHz repetition rate into a combination of optical fibers, yielding a spectrum that extends from 1547 nm (6464~cm$^{-1}$) to 3700~nm (2700~cm$^{-1}$) \cite{Amiot:17,mikkonen2018broadband}. The SC light is collimated into a beam of 2~mm diameter using a reflective fiber collimator, and then spectrally filtered. A long-pass filter with a cut-off at 2989~nm (3346~cm$^{-1}$, Northumbria Optical Coatings SLWP-2989) removes optical power from spectral regions not under interest and thus suppresses excess absorption from water (H$_2$O) molecules in the gas cell. As we have shown previously \cite{mikkonen2020noise}, eliminating unutilized absorption decreases the noise level significantly, as the intensity noise of the SC is coupled to the PA signal via light absorption by gas molecules. The filtered SC light with 21~mW of optical power is directed into a Fourier transform spectrometer (FTS, Bruker IRCube), whose maximum spectral resolution is 1~cm$^{-1}$. The optical path difference scanning velocity of the FTS is about 1~mm/s, which leads to modulation frequencies of 270--340~Hz at the wavenumber range of the filtered SC.

About 30\% of the incoming light is transmitted through the FTS and focused towards a commercial PA analyser (Gasera PA201) using a parabolic mirror (f = 20~cm). Within the PA analyser, a miniature non-resonant gas cell is 95~mm long, 4~mm in diameter, about 7~ml by total volume and sealed with anti-reflection coated BaF$_2$ windows (Thorlabs WG00530-E). Four mass flow controllers (Brooks Instruments 0154 and three pieces of Bronkhorst F-201CV) and a gas exchange system incorporated into the PA analyser control the gas flow into the cell with pressure and temperature set to 1~bar and 23~$^\text{o}$C, respectively.

The multipass configuration for this gas cell was designed through careful modelling of the optical system using Matlab. Specifically, we simulated the propagation of a Gaussian beam \cite{tovar1995generalized} in different Herriott configurations \cite{tarsitano2007multilaser,altmann1981two} by varying the properties of the mirrors (radius of curvature and separation) and the input beam (radius of curvature and beam waist). The optimum number of passes was found to be ten using mirrors with 200~mm radius of curvature and separated by 138~mm. Such gold-plated spherical mirrors (LBP Optics) with diameters of 12.7~mm were placed outside the gas cell, where the SC radiation is guided off-axially through a small aperture (1~mm in diameter, displaced 1.1~mm from the mirror center) in the front mirror. A simulated intensity pattern at the front mirror is shown at the bottom-right corner in Fig. \ref{fig:setup}. A theoretical signal enhancement (dashed black line in Fig. \ref{fig:methane}b) for this multipass system is slightly wavenumber dependent around 6.2, considering the transmittance curve of the windows (96\% on average) reported by the manufacturer and a constant 98\% mirror reflectance.

The cantilever microphone with an interferometric displacement detection is attached to the side of the gas cell, recording an interferogram containing information on all the excited acoustic waves, whose frequencies are directly proportional to specific absorbed wavenumbers. A three-term Blackman-Harris apodization function is applied for the resampled (reference HeNe laser) interferogram, and the absorption spectrum is obtained by Fourier transform.

\section{Results and discussion}

\subsection{Multipass enhancement and system performance}
\begin{figure}[h]
 \centering
 \includegraphics[height=7.7cm]{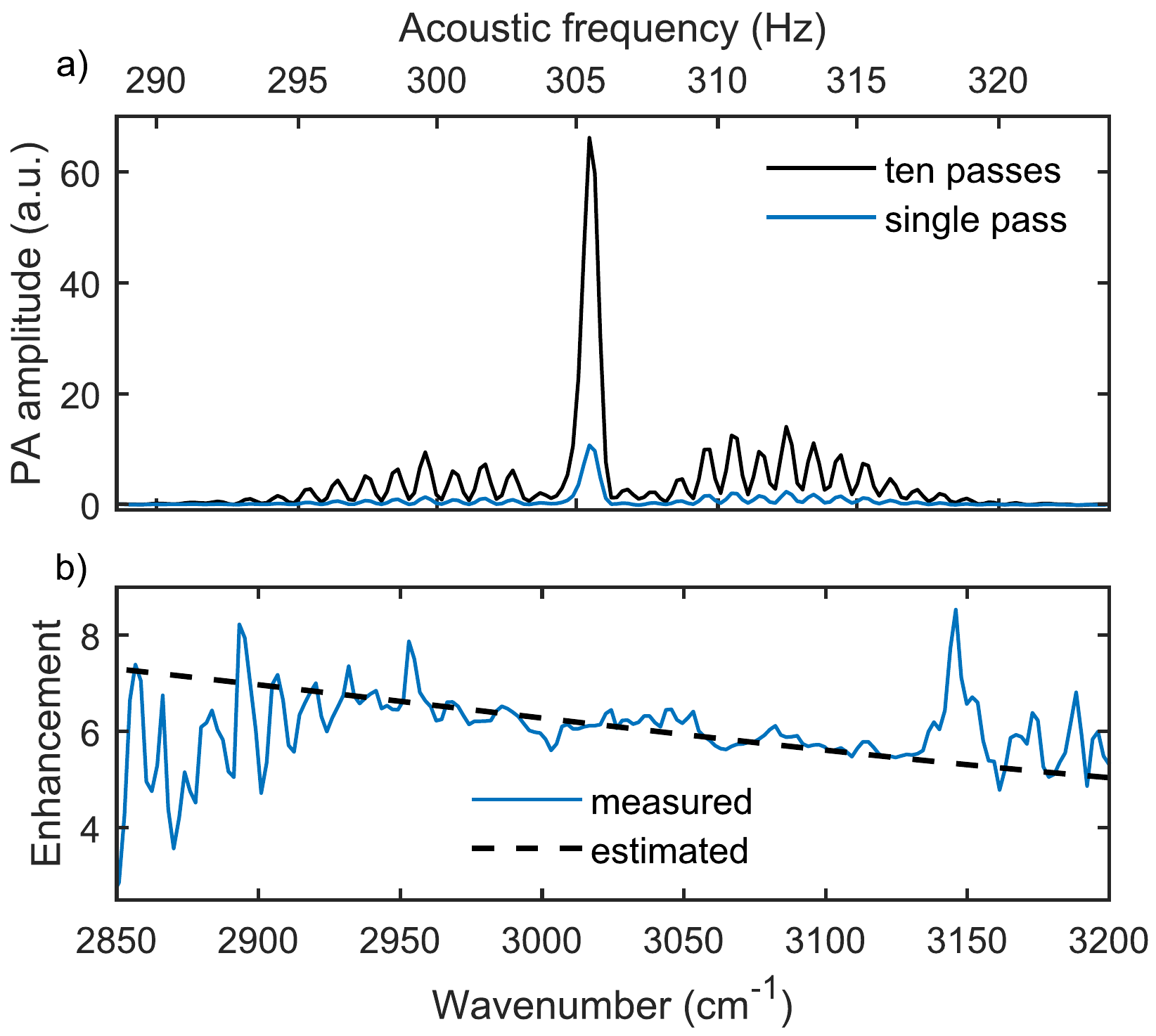}
 \caption{a) Measured FT-PAS spectra of 36~ppm CH$_4$ in N$_2$ using the multipass enhancement (black line) and allowing only one beam pass through the gas cell (blue line). Both spectra were averaged over 40~s (eight scans) at a spectral resolution of 4~cm$^{-1}$. In addition to the optical domain on the bottom, the corresponding modulation frequencies in the acoustic domain are shown on the top. b) Signal enhancement using the multipass arrangement, evaluated from the measured spectra (solid blue line) and estimated from the window transmittance and mirror reflectance (dashed black line).}
 \label{fig:methane}
\end{figure}

We measured the rovibrational absorption spectra of the fundamental C--H stretch band of methane (CH$_4$) around 3000~cm$^{-1}$ to demonstrate multipass enhancement and to compare the performance of the system with previous studies. Figure \ref{fig:methane} shows the absorption spectrum of 36 parts per million (ppm, volume mixing ratio) of CH$_4$ in nitrogen (N$_2$), measured both with a single and ten beam-pass through the gas cell. Single pass was realized by blocking the beam before the back spherical mirror. Each spectrum was averaged over eight scans at a spectral resolution of 4~cm$^{-1}$ resulting in a total averaging time of 40~s. A small H$_2$O residual was subtracted from both spectra using a methane-free spectral region with water absorption around 3200~cm$^{-1}$ and a separate pure H$_2$O measurement. We calculated the detection enhancement arising from the multipass configuration by comparing the raw spectra with and without the multipass enhancement. This division shown in Fig. \ref{fig:methane}b (solid blue line) is in excellent agreement with the theoretical prediction of approximately a sixfold signal enhancement, especially in a region with high signal-to-noise ratio.

For the multipass enhanced system, we estimated the limit of detection as LOD = 3$\sigma c$/$S$, where $c$ is the gas concentration (36~ppm), $S$ is the strongest spectral peak evaluated at 3017~cm$^{-1}$ and $\sigma$ is the noise level calculated from a separate 40~seconds long recording with only N$_2$ in the gas cell, as this represents the noise level at low CH$_4$ concentrations \cite{mikkonen2020noise}. Due to a small H$_2$O residual in the N$_2$ measurement, the noise level was calculated as the standard deviation of the baseline at 2200--2800~cm$^{-1}$. As will be shown later, this choice is justified by the flat noise level across the investigated wavenumber region. Based on this calculation, a detection limit of 32~ppb in 40~s for CH$_4$ was obtained. 

The performance of our system is compared to that of previous FT-PAS studies in Table \ref{tbl}. Note that the power spectral density is reported without any multipass enhancement (a double-pass arrangement was used in Refs. 8, 10 and 11). Significantly, our system shows a factor of 12 improvement in the time-normalized detection limit compared to the best previous FT-PAS system \cite{mikkonen2020noise} due to the multipass enhancement and a more sensitive PA analyser. In general, the differences between the LODs in Table \ref{tbl} also result from the power spectral densities, spectral resolutions and 
from the procedures for estimating the noise level at the detection limit. The LOD obtained here is comparable with other relatively simple broadband mid-IR techniques such as SC-based direct absorption spectroscopy and dual-comb spectroscopy with achieved CH$_4$ detection limits of 47~ppb and 2.4~ppm (for $\sim$100 times weaker absorption cross sections), respectively \cite{jahromi2020sensitive,sterczewski2019mid}. However, both of these techniques require two orders of magnitude larger sample volumes compared to our approach.

\begin{table}[h]
\centering
  \caption{\ A summary of recent FT-PAS experiments using supercontinuum (SC) or optical frequency comb (OFC) light sources in terms of CH$_4$ limit of detection (LOD, 3$\sigma$, 100~s), spectral resolution and incident power spectral density (PSD).}
  \label{tbl}
  \begin{tabular*}{0.7\textwidth}{@{\extracolsep{\fill}}lllll}
    \hline
    Source & LOD (ppb) & Res. (cm$^{-1}$) & PSD (\textmu W/cm$^{-1}$) & Ref. \\
    \hline
    SC  & \phantom{0}990  & \phantom{0}4     & \phantom{0}31 & 8 \\
    OFC & 3300 & \phantom{0}0.033 & \phantom{0}21 & 10 \\ 
    OFC & \phantom{0}660  & \phantom{0}0.02  & 240 & 11 \\ 
    SC  & \phantom{0}240  & 13 & \phantom{0}19 & 9 \\ 
    SC  & \phantom{00}20   & \phantom{0}4    & \phantom{0}18 & This work \\
    \hline
  \end{tabular*}
\end{table}

\begin{figure}[h]
 \centering
 \includegraphics[height=6cm]{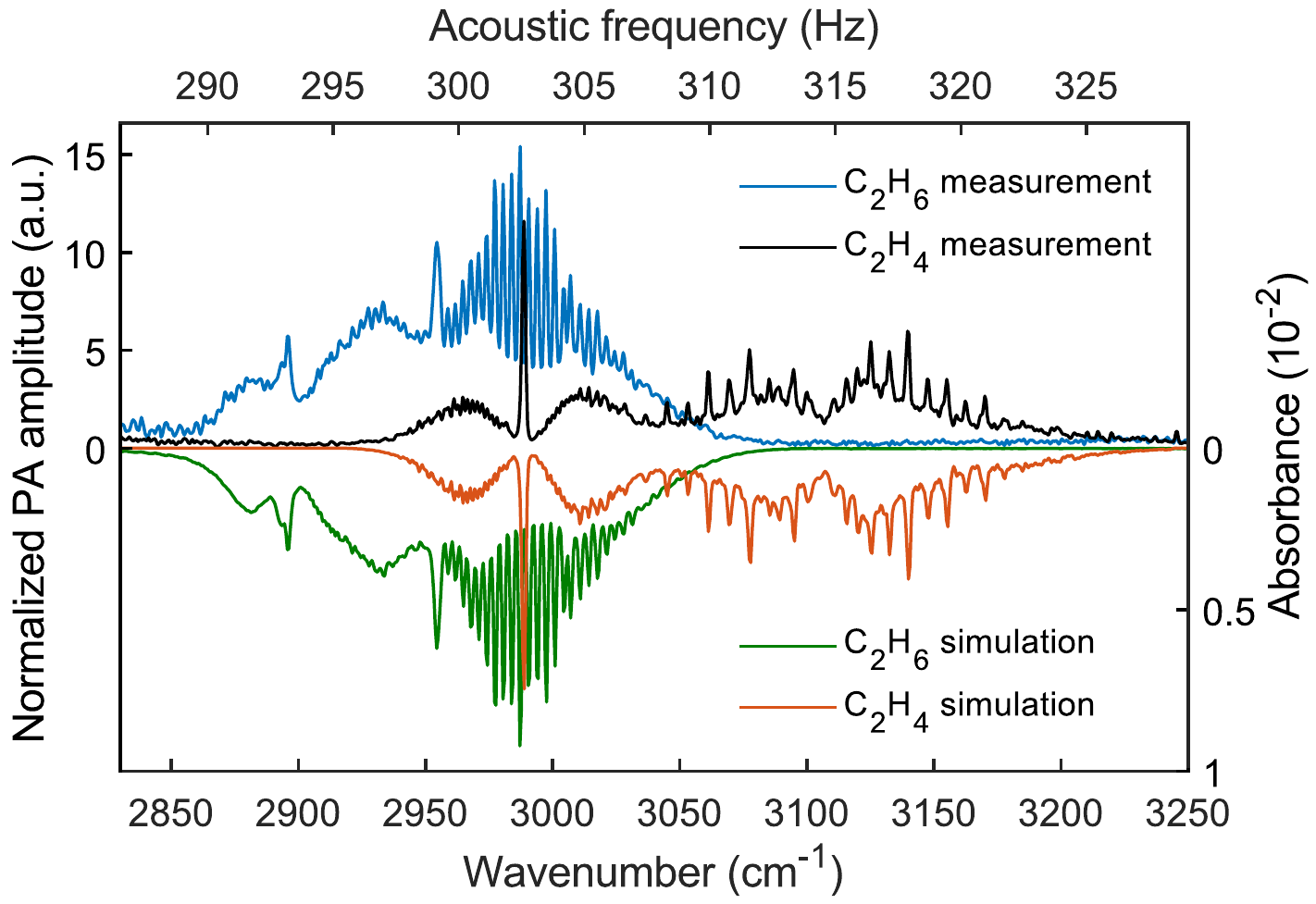}
 \caption{Measured spectra of 30~ppm C$_2$H$_6$ and 100~ppm C$_2$H$_4$ at 1~cm$^{-1}$ spectral resolution, compared to the simulated absorbance based on the HITRAN database (inverted). A small H$_2$O residual was subtracted from the raw spectra, which were also normalized by the envelope of the SC spectrum.}
 \label{fig:ethene}
\end{figure}

The performance of the system was further characterized by measuring the C--H stretch bands of 30~ppm ethane (C$_2$H$_6$) and 100~ppm ethene (C$_2$H$_4$) separately at 1~cm$^{-1}$ spectral resolution. The raw spectra were averaged over 200~s (ten scans), corrected for H$_2$O absorption using the procedure described for CH$_4$ and normalized by the spectral envelope of the SC measured with a monochromator (Spectral Products DK480 1/2) and numerically corrected by the transmittance of the filter and the windows. These corrected spectra, shown in Fig. \ref{fig:ethene}, are in excellent agreement with simulations based on the HITRAN database \cite{hill2016hitranonline} and the instrument lineshape function. The increase in the baseline at the low wavenumber part of the C$_2$H$_6$ spectrum is due to low power spectral density of the SC in this region, which amplifies the background noise in the normalization process.

\subsection{Calibration and noise characterization}
\begin{figure}[h]
 \centering
 \includegraphics[height=11cm]{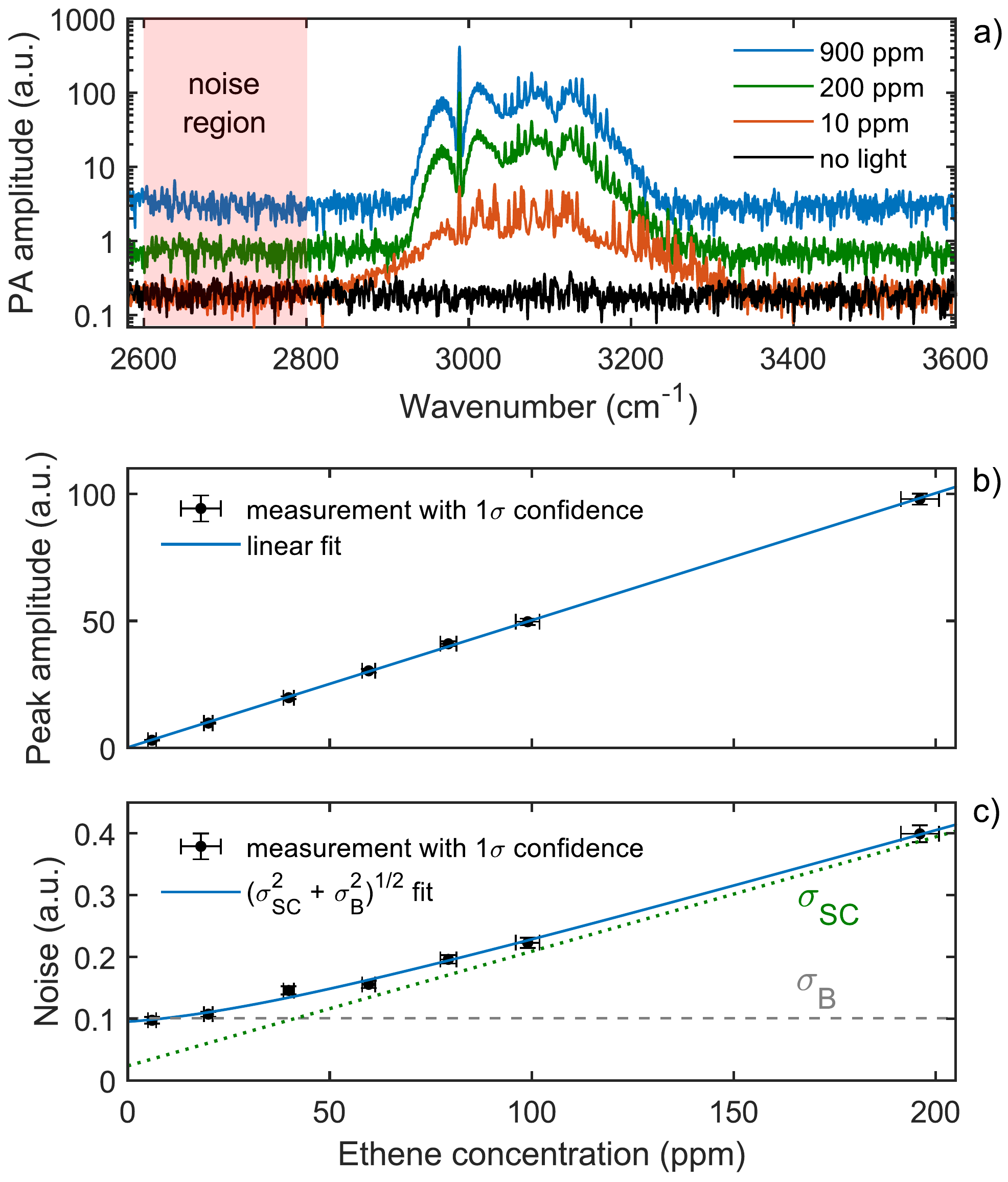}
 \caption{a) Measured absorption spectra of C$_2$H$_4$ at three concentrations and a background spectrum (no light entering the cell), all averaged over 100~s (5~scans) at a spectral resolution of 1~cm$^{-1}$. The spectra of C$_2$H$_4$ include H$_2$O residual, which is especially visible at the lowest concentration. Red area shows the noise region used in estimating the detection limits, which is justified by the flat noise level across the wavenumber region. b) Peak amplitude and c) noise level with respect to the applied C$_2$H$_4$ concentration. Noise has two contributions: Brownian noise ($\sigma_{\text{B}}$) from the thermal fluctuations of gas molecules corresponding to the background spectrum in a), and a concentration dependent component from the light source ($\sigma_{\text{SC}}$).}
 \label{fig:linearity}
\end{figure}

We studied the effect of the gas concentration on the amplitude of the measured photoacoustic signal and the noise level. The concentration--amplitude dependence is used for calibration, which is always required in PAS, and the noise relation gives insight into the unique noise mechanism of FT-PAS studied in more detail elsewhere \cite{mikkonen2020noise}. Samples of C$_2$H$_4$ with varying concentration between 6 and 900~ppm were prepared by diluting C$_2$H$_4$ from a 1\% gas bottle to N$_2$. The measured spectra for the different concentrations with 1~cm$^{-1}$ resolution and averaged over 100~s (5~scans) are shown in Fig. \ref{fig:linearity} a on a logarithmic scale. One can see how both the signal amplitude and noise level increases with the concentration. We also observe that the noise level remains constant across the characterized wavenumber range, justifying the use of the noise value in a specific region (red area in Fig. \ref{fig:linearity}a) to assess the detection limit. For each spectrum, the amplitude was calculated from the strongest spectral peak of C$_2$H$_4$ after H$_2$O subtraction, and the noise level was estimated as the standard deviation of the raw PA amplitude in an absorption-free region at 2600--2800~cm$^{-1}$. 

Figures \ref{fig:linearity}b,c illustrate quantitatively the dependence of the signal amplitude and noise as a function of C$_2$H$_4$ concentration up to 200~ppm. Uncertainties in the C$_2$H$_4$ concentrations result from dilution errors caused by the gas bottle uncertainty (2\%) and the errors in the mass flow controllers (1--17\% depending on the concentration). Uncertainties in the peak amplitudes and the noise levels are mainly caused by the fluctuation in the SC power, which was estimated from a long-term stability measurement. As can be expected, the amplitude of the acoustic signal scales linearly with the concentration (R$^2$=0.9997). At C$_2$H$_4$ concentrations beyond 200~ppm, the amplitude begins to saturate in accordance with the Beer-Lambert law. This observed amplitude nonlinearity at high concentrations was compared to a theoretical nonlinearity of the C$_2$H$_4$ amplitude to retrieve the effective optical path length. This simulation, based on a reference spectrum from the HITRAN database \cite{ROTHMAN20134}, the Beer-Lambert law and the instrument lineshape function, yields an optical path length of 58$\pm$10~cm corresponding to a path length enhancement of 6.1$\pm$1.1 and matching well to our achieved signal enhancement in the CH$_4$ demonstration. 

The noise level exhibits similar dependence with the C$_2$H$_4$ concentration as the amplitude, because the intensity noise of the SC couples to the PA signal via light absorption by gas molecules. Higher concentration yields more absorption and therefore both higher amplitude and noise. This concentration dependent noise component is shown as a dotted green line in Fig. \ref{fig:linearity}c, extracted from the fit. However, the noise has another contribution from the thermal fluctuations of gas molecules (Brownian noise), which begins to dominate below 50~ppm, causing the noise level to approach the fundamental limit. This background noise level (the dashed grey line in Fig. \ref{fig:linearity}c) was calculated from the measurement with no light entering the cell (the black line in Fig. \ref{fig:linearity}a) and is consistent with the fitting result. As the noise at low concentrations equals the fundamental noise level, light absorption by the cell windows is expected to be small, which supports our choice for the external multipass system.

\subsection{Long-term stability}
\begin{figure}[h]
 \centering
 \includegraphics[height=9cm]{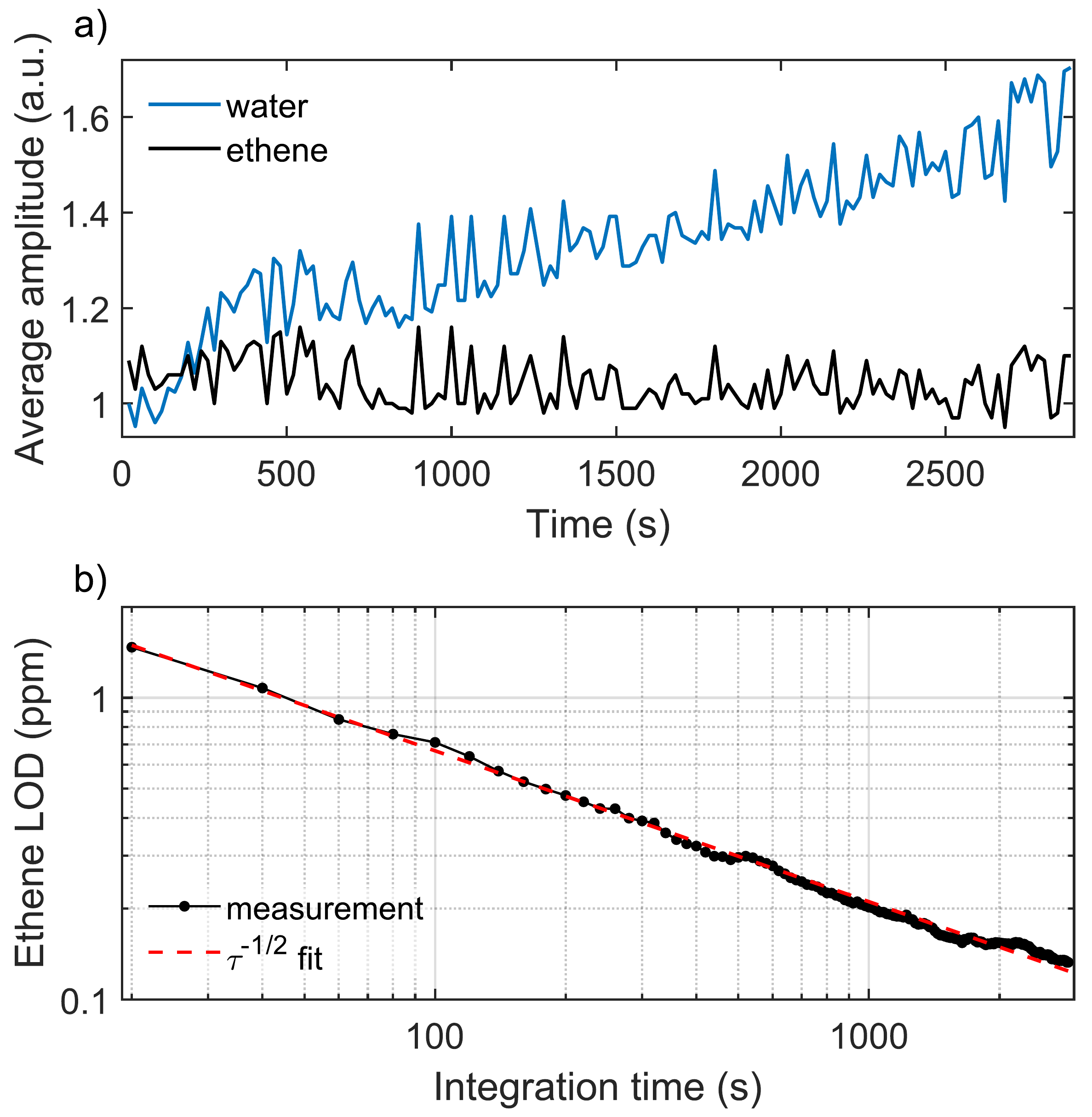}
 \caption{a) Separated average amplitudes of H$_2$O and C$_2$H$_4$ during the stability measurement, in which a 20~s scan was recorded every 40~s for 5760~s. Due to this dead time, in reality the drifts and variations in the amplitudes occur twice as slow as shown in the plot. b) Calculated limit of detection (LOD) of C$_2$H$_4$ for different effective integration times. Water absorption was subtracted from each spectrum to account for the increase of H$_2$O concentration during the measurement.}
 \label{fig:Allan}
\end{figure}

To evaluate the long-term stability of the SC-based FT-PAS system, we measured 100~ppm of C$_2$H$_4$ every 40~s for 8320~s. Due to the limitations in the FTS, each 20~s scan followed by a 20~s break, cutting the averaging time in half. Furthermore, distorted measurements (30\%) due to external vibrations were removed using Hampel's outlier test for average amplitudes in the frequency range of 710--870~Hz (7030--8610~cm$^{-1}$) resulting in an effective averaging time of 2880~s. Water was again subtracted from each spectrum by fitting pure H$_2$O and C$_2$H$_4$ spectra to each raw spectrum. The result of this fitting routine is shown in Fig. \ref{fig:Allan}a, which displays the separated water and ethene traces during the measurement.
A significant increase in the H$_2$O amplitude is due to desorption from the cell walls. The C$_2$H$_4$ amplitude, on the other hand, stays constant on average but exhibits variations most likely due to SC instability. Corresponding variations are also visible in the H$_2$O trace. The relative precision of the system (4.8\%) was estimated from the standard deviation of the ethene trace relative to its mean. The minimum precision of the system (at 1~cm$^{-1}$ spectral resolution and for a single scan) is 200~ppb, set by the Brownian noise.

Using a water corrected ethene amplitude at 2980~cm$^{-1}$ and noise from the raw spectrum (estimated as described previously), we calculated C$_2$H$_4$ LODs for various integration times. The result of this averaging process shown in Fig. \ref{fig:Allan}b displays high stability of the system and a detection limit of 130~ppb in 2880~s for C$_2$H$_4$. However, as previously discussed, the correct noise level for the detection limit calculation should be from the N$_2$ measurement, and thus a practical C$_2$H$_4$ detection limit is expected to be about two times smaller. The curve in Fig. \ref{fig:Allan}b follows the inverse square root law well on average, as the main noise contributions (pulse-to-pulse SC fluctuations and Brownian noise) exhibit white noise characteristics in this spectral region. Deviations from the theoretical line are caused by the long-term variations of the SC power spectral density and the increase of H$_2$O inside the cell, which raises the noise level during the measurement.

\subsection{Multi-species detection}
\begin{figure}[ht!]
 \centering
 \includegraphics[height=9.5cm]{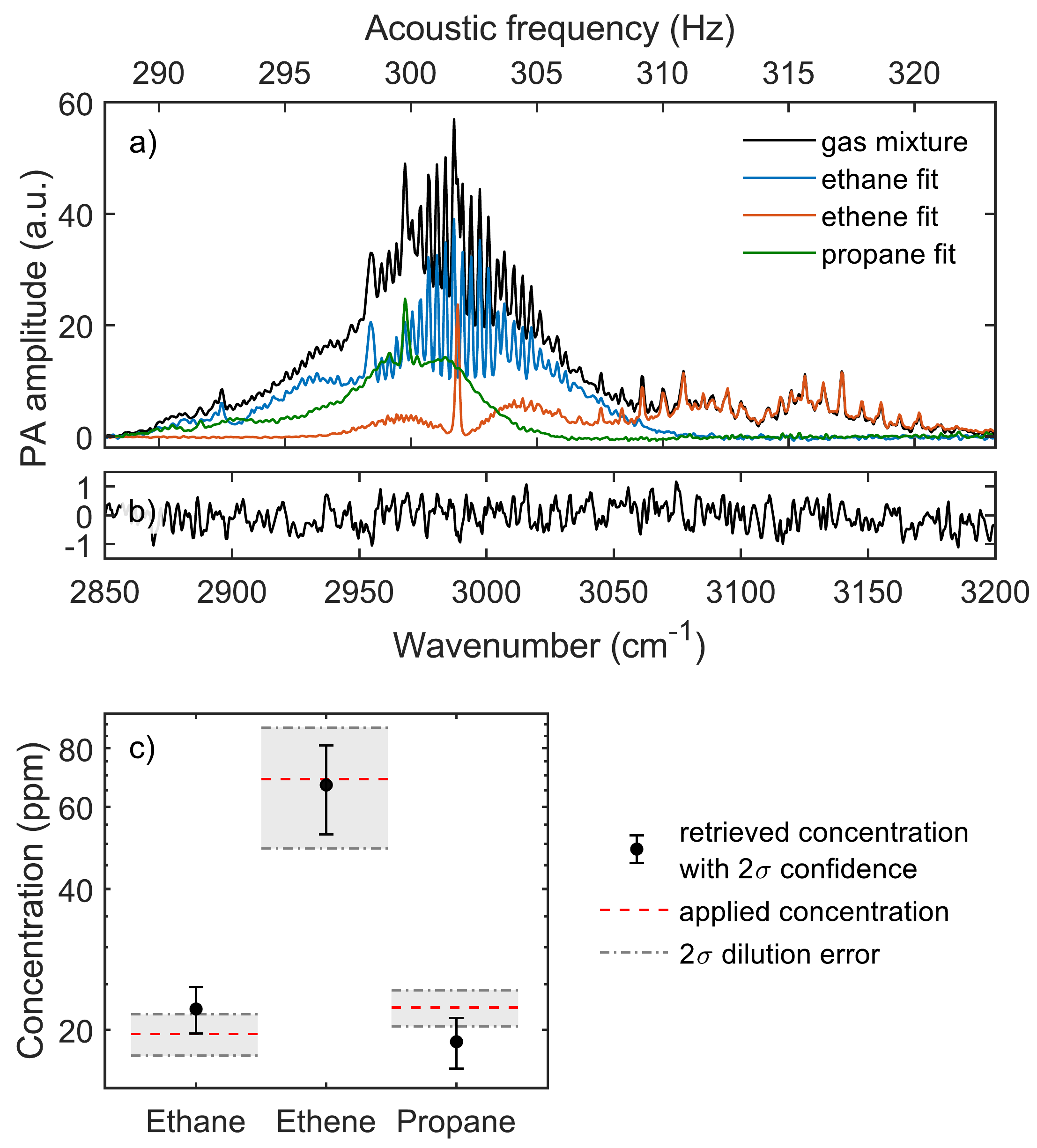}
 \caption{a) Measured FT-PAS spectra of a gas mixture (black) and individual species (blue, red and green), whose magnitudes were retrieved from a simple least-squares fitting routine. b) Residual of the spectral fitting. c)~Concentrations retrieved from the gas mixture (black dots), in comparison with the applied concentrations (red dashed lines).}
 \label{fig:mixture}
\end{figure}

We demonstrated the advantage of the high system linearity and spectral resolution with the wide spectral bandwidth in retrieving the concentrations of three species from a complex gas mixture. We first measured the reference spectra of ethane (30~ppm), ethene (100~ppm) and propane (20~ppm) by diluting each from a 1\% gas bottle to N$_2$. We then prepared a gas mixture containing 20~ppm ethane, 69~ppm ethene and 23~ppm propane, injected the mix to the gas cell and measured the spectrum. Water residual was removed from all spectra. A simple least-squares spectral fitting routine was used to match the weighted reference spectra to the spectrum of the gas mixture. The resulting scaled spectra of the individual species and the measured gas mixture spectrum are shown in Fig. \ref{fig:mixture}a and the residual of the spectral fit in Fig. \ref{fig:mixture}b.

The concentration of each species in the gas mixture was calculated by multiplying the concentration of the reference measurement with the optimum weight from the fitting routine. Fig.~\ref{fig:mixture}c illustrates the correspondence between the applied and the retrieved concentrations and shows that the linear spectral fitting is able to retrieve concentrations within the measurement uncertainty. Uncertainties in the applied concentrations result from the dilutions errors caused by the four mass flow controllers, which produce uncertainty also in the retrieved concentrations (via reference measurements) together with the system precision estimated previously.

\section{Conclusions}

A broadband mid-infrared optical gas sensing technique, Fourier transform photoacoustic spectroscopy (FT-PAS) with an external and compact Herriott-type multipass configuration has been demonstrated. Our FT-PAS system incorporates a home-built supercontinuum (SC) source with a high power spectral density and a cantilever enhanced photoacoustic analyser with a small sample volume. We characterized the fundamental C--H stretch bands of four hydrocarbons, methane, ethane, ethene and propane in nitrogen, at spectral resolutions of 1~cm$^{-1}$ and 4~cm$^{-1}$ with an excellent agreement with simulated absorption spectra. The Herriott cell for ten beam passes provides a sixfold broadband signal enhancement, resulting in a detection limit of 32~ppb in 40~s for methane. This is a factor of 12 lower compared to the best previous FT-PAS demonstrations and on the same scale with other broadband optical sensors with significantly higher gas consumption \cite{jahromi2020sensitive,sterczewski2019mid}, when normalized by the averaging time. Moreover, we demonstrated high linearity and good stability of the system, which was shown to provide a simple and accurate spectral analysis of complex gas mixtures.

The detection limit of SC-based FT-PAS can be improved with more transparent windows and further optimization of the multipass arrangement. Ideally, the PA analyser should be redesigned with the multipass enhancement in mind, as increasing the diameter of the gas cell allows more beam passes but decreases the sensitivity of PA detection. The detection limit can be further lowered with commercial SC sources, which currently provide three times higher power spectral density around 3000~cm$^{-1}$ compared to our device. Such sources would also enhance the precision of the system, which can be further improved by recording the optical power of the exiting radiation for each scan. As this light has propagated through a scanning Fourier transform spectrometer, an entire SC spectrum could also be continuously acquired for better spectral correction.

\section*{Conflicts of interest}
There are no conflicts of interest to declare.

\section*{Acknowledgements}
This work was supported by the Graduate School of Tampere University and the Academy of Finland Flagship Programme PREIN (320165).



\printbibliography

\end{document}